\documentclass{aa}  

\usepackage{graphicx}
\usepackage{txfonts}
\usepackage{gensymb}
\usepackage{wasysym}
\usepackage{caption}
\usepackage{subcaption}
\usepackage{threeparttable}
\usepackage{pdflscape}
\usepackage{rotating}
\usepackage[normalem]{ulem}

\usepackage{natbib}
\bibpunct{(}{)}{;}{a}{}{,}

\usepackage[dvipsnames]{xcolor}
\usepackage{hyperref}
\hypersetup{
    colorlinks=true,
    linkcolor=MidnightBlue,
    filecolor=blue,      
    urlcolor=cyan,
    citecolor=TealBlue
}

\begin{document}

   \title{TANAMI: Tracking Active Galactic Nuclei with Austral
Milliarcsecond Interferometry}

   \subtitle{III. First-epoch S band images}

   \author{P. Benke
          \inst{1,2}
          \and
          F. R\"osch\inst{2}
          \and
          E. Ros\inst{1}
          \and
          M. Kadler\inst{2}
          \and
          R. Ojha\inst{3}
          \and
          P. G. Edwards\inst{4}
          \and
          S. Horiuchi\inst{5}
          \and
          L.~J. Hyland\inst{6}
          \and
          C. Phillips\inst{4}
          \and
          J.~F.~H. Quick\inst{7}
          \and
          J. Stevens\inst{4}
          \and
          A.~K. Tzioumis\inst{4}
          \and
          S. Weston\inst{8}
          }

   \authorrunning{P. Benke et al.}

   \institute{Max Planck Institute for Radio Astronomy, Auf dem H\"ugel 69, D--53121 Bonn, Germany
         \and
            Julius Maximilians University W\"urzburg, Faculty of Physics and Astronomy, Institute for Theoretical Physics and Astrophysics, Chair of Astronomy, Emil-Fischer-Str. 31, D-97074 Würzburg, Germany
        \and
            NASA HQ, 300 E St SW, DC 20546-0002, Washington DC, USA
        \and
            CSIRO Space and Astronomy, PO Box 76, Epping, NSW 1710, Australia
            \and
            CSIRO Space and Astronomy, Canberra Deep Space Communications Complex, PO Box 1035, Tuggeranong, ACT 2901, Australia
            \and
            School of Natural Sciences, University of Tasmania, Private Bag 37, Hobart, Tasmania 7001, Australia
            \and
             Hartebeesthoek Radio Astronomy Observatory, PO Box 443, 1740 Krugersdorp, South Africa
            \and
            Space Operations New Zealand Ltd, Hargest House, PO Box 1306, Invercargill 9840, New Zealand
             }

   \date{Received ; accepted}

  \abstract{With the emergence of very high energy astronomy (VHE; $E>100$~GeV), new open questions were presented to astronomers studying the multi-wavelength emission from blazars. Answers to these open questions, such as the Doppler crisis, and finding the location of the high-energy activity have eluded us thus far. Recently, quasi-simultaneous multi-wavelength monitoring programs have shown considerable success in investigating blazar activity.}
  {After the launch of the \textit{Fermi Gamma-ray Space Telescope} in $2008$, such quasi-simultaneous observations across the electromagnetic spectrum became possible. In addition, with very long baseline interferometry (VLBI) observations we can resolve the central parsec region of active galactic nuclei (AGN) and compare morphological changes to the $\gamma$-ray activity to study high-energy emitting blazars. To achieve our goals, we need sensitive, long-term VLBI monitoring of a complete sample of VHE detected AGN.}
  {We performed VLBI observations of TeV-detected AGN and high likelihood neutrino associations as of December of 2021 with the Long Baseline Array (LBA) and other southern hemisphere radio telescopes at $2.3$\,GHz.}
  {In this paper we present first light TANAMI S-band images, focusing on the TeV-detected sub-sample of the full TANAMI sample. Apart from these very high energy-detected sources, we also show images of the two flux density calibrators and two additional sources included in the observations. We study the redshift, $0.1-100$~GeV photon flux and S-band core brightness temperature distributions of the TeV-detected objects, and find that flat spectrum radio quasars and low synchrotron peaked sources on average show higher brightness temperatures than high-synchrotron-peaked BL\,Lacs. Sources with bright GeV gamma-ray emission also show higher brightness temperature values than $\gamma$-low sources.}
  {Long-term monitoring programs are crucial to study the multi-wavelength properties of active galactic nuclei. With the successful detection of even the faintest sources with flux densities below $50$~mJy, future work will entail kinematic analysis and spectral studies both at $2.3$ and $8.4$\,GHz to investigate the connection between the radio and $\gamma$-ray activity of these objects.}


   \maketitle

\section{Introduction}

Blazars are active galactic nuclei (AGN) whose jets are oriented towards the line of sight of the observer. As a result of this, their emission is highly beamed, and they often exhibit apparent superluminal jet motion. Their two-humped spectral energy distribution (SED) can be modeled with a lower energy, synchrotron component and a high-energy component arising due to leptonic \citep{1992ApJ...397L...5M} and/or hadronic processes \citep{1993A&A...269...67M}. Blazars can be further classified as BL\,Lac objects and flat spectrum radio quasars (FSRQ). The former can be divided into four sub-classes, low (LBL, $\nu_{\mathrm{peak}}<10^{14}$~Hz), intermediate (IBL, $10^{14}<\nu_{\mathrm{peak}}<10^{15}$~Hz), high (HBL, $10^{15}<\nu_{\mathrm{peak}}<10^{17}$~Hz), and extremely high (EHBL, $\nu_{\mathrm{peak}}>10^{17}$~Hz) synchrotron peaked objects, based on the location of the synchrotron peak, $\nu_{\mathrm{peak}}$, in their SED \citep{2010ApJ...716...30A}. However, these objects still present us with many open questions, including the launching mechanisms responsible for the creation of jets, the collimation and acceleration of jet material, as well as the origins of the multi-wavelength emission. 

\begin{figure*}[h!]
    \centering
    \begin{subfigure}{0.45\linewidth}
        \includegraphics[width=\linewidth, angle=-90]{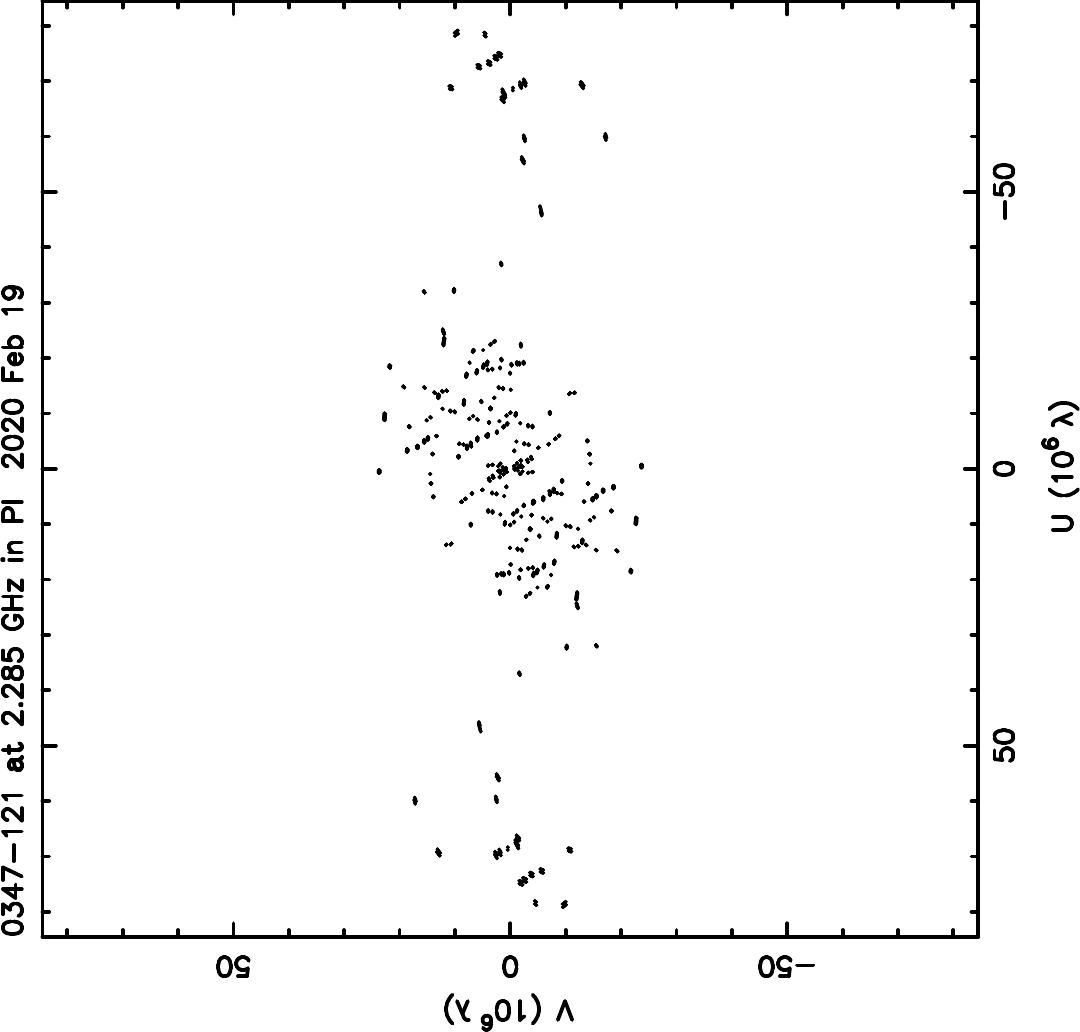}
    \end{subfigure}
    \begin{subfigure}{0.45\linewidth}
        \includegraphics[width=\linewidth, angle=-90]{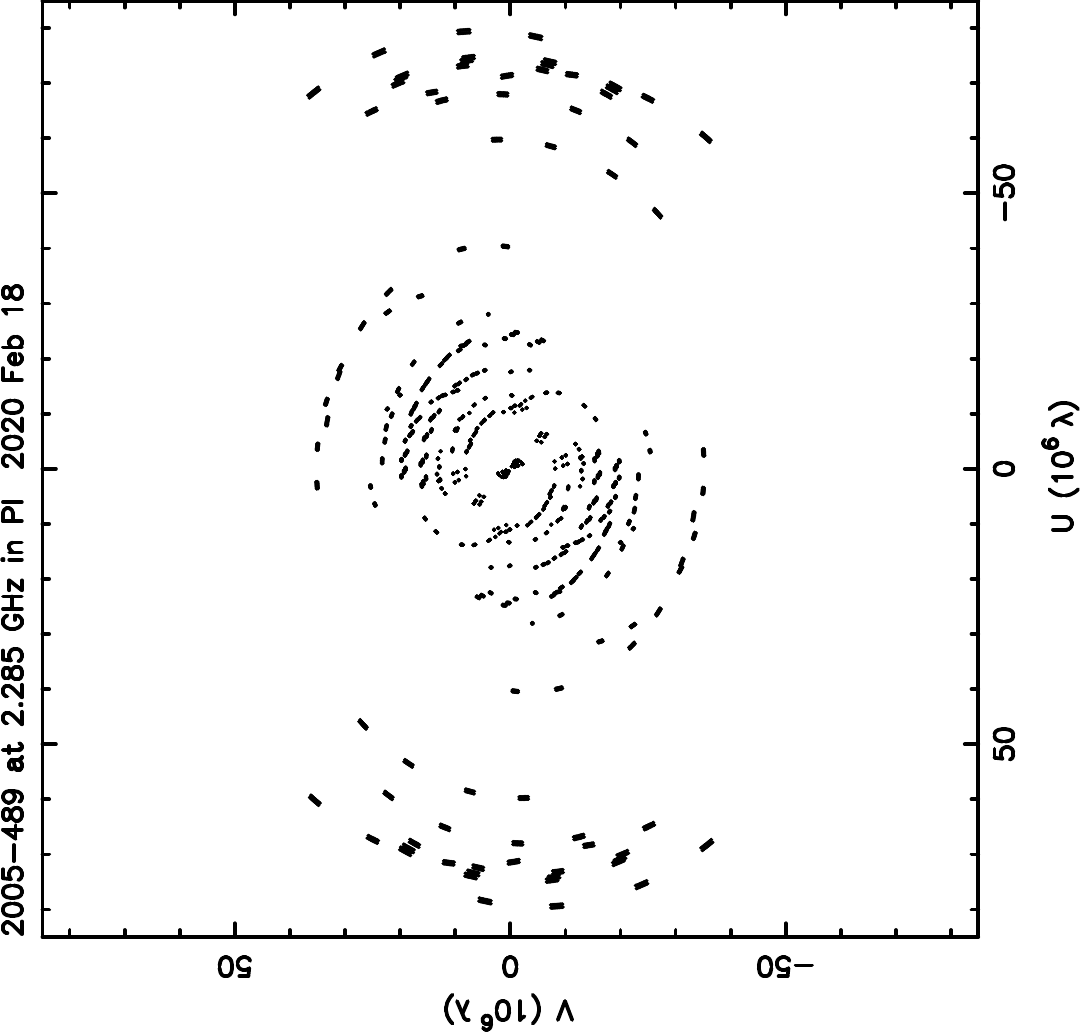}
    \end{subfigure}
    \caption{TANAMI S band ($u$,$v$) coverages in the most extended array configuration for low and high declination sources. Due to the geographical layout of the array at the southern hemisphere we do not have antenna pairs that cover intermediate-length baselines.}
    \label{fig:uv}
\end{figure*}

One of the widely studied open questions regarding the multi-wavelength nature of blazars is known as the Doppler crisis \citep{2018mgm..conf.3074P}. Many blazars have been detected at high (HE; $100$~MeV $<E<300$~GeV) and very high energies (VHE; $E>100$~GeV), and they show rapid variability, often on timescales of minutes \citep[see e.g., 2155$-$304][]{2007ApJ...664L..71A}. This suggests a small emission region and high Doppler factors, and indeed, Doppler factors of $\delta\approx50$ are required to reproduce the spectral energy distribution (SED) at high energies \citep{2018mgm..conf.3074P}. On the other hand, very long baseline interferometry (VLBI) observations reveal slow component motions with Doppler factors $<10$ \citep{2018mgm..conf.3074P}. A prominent example of the Doppler crisis blazar is Mrk\,$421$ \citep{mrk421tev, lico12}. 
Locating the origin of the high-energy emission, also known as the blazar zone, has also proven to be challenging. The observed short-time variability in the VHE band \citep[see for example 2155$-$304][]{2010A&A...520A..23R} suggests that the TeV emission originates from a small region. Under the assumption that the emission region fills the jet diameter, the VHE emission region must be located close to the central engine \citep{2013ApJ...766L..11S}. The availability of seed photons for external Compton scattering also supports this scenario \citep{2016ApJ...821..102B}. However, this high-energy emission is expected to be absorbed by the dense photon fields of the broad line region, so we could not detect blazars at VHE. On the other hand, \citet{2012arXiv1204.6707M} and \citet{2013ApJ...773..147J} found that activity near the radio core at $43$\,GHz coincides with $\gamma$-ray flares, suggesting that the high energy emission originates downstream of the central supermassive black hole.

Several models have been proposed to explain both the Doppler crisis and the location of the blazar zone, all of them invoking multiple Doppler factors for different emission processes in the parsec-scale jet. These models include a spine-sheath transverse velocity structure \citep{2005A&A...432..401G}, a decelerating jet with a slower moving plasma at the jet edge \citep{2003ApJ...594L..27G}, multi-zone models \citep{2011A&A...534A..86T}, or minijets created via magnetic reconnection \citep{2009MNRAS.395L..29G}. Currently it is only possible to distinguish between these jet models with high resolution VLBI observations.

\begin{table*}[h]
\centering
\caption{Antennas participating in the observations.}
\label{tab:antennas}
    \begin{tabular}{l  c c l c c }
    \hline\hline
        Telescope & Abbreviation & Diameter (m) & Location & Latitude & Longitude\\
        \hline
        Parkes & Pa & $64$ & Parkes, New South Wales, Australia & $32\degr59\arcmin52\arcsec$ S& $148\degr15\arcmin47\arcsec$ E\\
        ATCA & At & $5\times22$ & Narrabri, New South Wales, Australia & $30\degr18\arcmin46\arcsec$ S & $149\degr33\arcmin01\arcsec$ E \\
        Mopra & Mp & $22$ & Coonabarabran, New South Wales, Australia & $31\degr16\arcmin04\arcsec$ S & $149\degr06\arcmin00\arcsec$ E \\
        Hobart & Ho & $26$ & Mt. Pleasant, Tasmania, Australia & $42\degr48\arcmin13\arcsec$ S & $147\degr26\arcmin26\arcsec$ E \\
        Ceduna & Cd & $30$ & Ceduna, South Australia & $31\degr52\arcmin04\arcsec$ S & $133\degr48\arcmin34\arcsec$ E\\
        DSS43\tablefootmark{a} & Ti & $70$ & Tidbinbilla, Australia & $35\degr24\arcmin05\arcsec$ S & $148\degr58\arcmin54\arcsec$ E \\
        DSS36\tablefootmark{a} & Td & $34$ & Tidbinbilla, Australia &$35\degr24\arcmin05\arcsec$ S & $148\degr58\arcmin54\arcsec$ E\\
        Hartebeesthoek & Hh & $26$ & Gauteng, South Africa & $25\degr53\arcmin25\arcsec$ S & $27\degr41\arcmin08\arcsec$ E \\
        Warkworth\tablefootmark{b, c} & Ww & $12$ & Auckland, New Zealand & $36\degr26\arcmin00\arcsec$ S & $174\degr39\arcmin46\arcsec$ E \\
        Katherine\tablefootmark{c} & Ke & $12$ & Northern Territory, Australia & $14\degr22\arcmin32\arcsec$ S & $132\degr09\arcmin09\arcsec$ E\\
        Yarragadee\tablefootmark{c} & Yg & $12$ & Western Australia & $29\degr02\arcmin50\arcsec$ S & $115\degr20\arcmin44\arcsec$ E \\
        \hline
    \end{tabular}
    \tablefoot{
      \tablefoottext{a}{Operated by the Deep Space Network of the USA National Aeronautics and Space Administration.}
      \tablefoottext{b}{Operated by the Space Operations New Zealand Ltd. \citep{2010ivs..conf..113G}.}
       \tablefoottext{c}{International VLBI Service for Geodesy and Astrometry (IVS) station.}
    }
\end{table*}

\vspace{0.5cm}
\begin{table*}[h]
    \centering
        \caption{Summary of observing sessions and participating antennas.}
    \label{tab:observations}
    \begin{tabular}{l c l}
    \hline\hline
        Epoch & Participating telescopes\tablefootmark{a} & Remarks \\
        \hline
        2020-02-18 & Pa, At, Mp, Ho, Cd, Hh, Ke, Yg, Td, Ti, Ww & Td and Ti recorded single polarization at a time \\
        2020-11-21 & Pa, At, Mp, Ho, Cd, Hh, Yg, Ww & Cd only in the first half of the observations due to recording problems \\
        2021-12-04 & Pa, At, Mp, Hb, Cd, Hh, Ke, Yg, Ww & Yg experienced recording problems \\
     \hline 
    \end{tabular}
        \tablefoot{
      \tablefoottext{a}{Telescopes are denoted as Parkes (Pa), ATCA (At), Mopra (Mp), Hobart (Ho), Ceduna (Cd), Hartebeesthoek (Hh), Katherine (Ke), Tidbinbilla $34$~m (Td), Tidbinbilla $70$~m (Ti), Warkworth (Ww), Yarragadee (Yg). See antenna characteristics in Tab.~\ref{tab:antennas}.}
    }
\end{table*}

To resolve these open questions we need long-term quasi-simultaneous multi-wavelength monitoring with high cadence. With the launching of \textit{Fermi} in 2008 \citep{2009ApJ...697.1071A}, continuous GeV $\gamma$-ray monitoring became possible with the Large Area Telescope instrument on board of the satellite. However, since high-energy telescopes lack the resolution to determine which part of the AGN is responsible for the $\gamma$-ray emission, we need additional high-resolution observations that can reveal the parsec scale structure of these objects. Tracking Active Galactic Nuclei with Austral Milliarcsecond Interferometry (TANAMI) is a VLBI monitoring program aiming at long-term observations of the rarely exploited southern sky. TANAMI started X and K band ($8.4$ and $22$\,GHz, respectively) observations in 2007 \citep{ojha10}. The program uses telescopes of the extended Long Baseline Array (LBA, see Tab.~\ref{tab:antennas}). Monitoring programs such as TANAMI, MOJAVE and the Boston University Blazar Group are needed to capture the different emission states of AGN and reveal their multiwavelength properties via tracking morphological and brightness changes of the targets. With the emergence of neutrino astronomy, the TANAMI sample has been expanded to include neutrino associations, and in 2020 S band observations were started to study faint TeV sources and extended jet structures. Currently, TANAMI is the sole AGN monitoring program focusing on southern-hemisphere sources, and our multi-frequency monitoring in the S and X bands (K-band observations were discontinued in 2020) will enable us to carry out spectral studies, coreshift measurements, etc., in the future.

In this paper we present the first light S band ($2.3$\,GHz) images of sources observed during the first three epochs in 2020-2021, focusing on the TeV-detected sub-sample of TANAMI. In Sect.~\ref{obs} we describe the sample selection, observations and data reduction, and in Sect.~\ref{results} we discuss the data analysis. In Sect.~\ref{sources} and Sect.~\ref{disc} we describe the clean images and properties of the sample, and finally we summarize our results in Sect.~\ref{summ}.

\section{The source sample, observations and data reduction}
\label{obs}

The TANAMI AGN sample currently consist of $183$ sources below the J2000 declination of $0$\degree. The original sample was defined as a radio and $\gamma$-ray selected sample below $-30$\degree declination in the X and K band \citep{ojha10}. However, with the emergence of neutrino astronomy, the declination range of the observations were broadened, and the sample was expanded to accommodate new astrophysical neutrino associations, as well as newly discovered TeV emitting AGN on the southern hemisphere. In $2020$, S band observations were introduced to TANAMI to study extended jet structures and to monitor faint ($S_{\mathrm{2.3 GHz}}<50$~mJy) sources that were previously excluded from the sample. In this paper we discuss our first light S band results on our TeV sample, which at the time of the observations (December, 2021) contained all known VHE emitter AGN on the southern celestial hemisphere.

Observations were carried out in S band with the ad hoc TANAMI array, utilizing Parkes, ATCA, Mopra, Hartebeesthoek, Ceduna, Hobart, Tidbinbilla and IVS stations, such as Katherine, Warkworth and Yarragadee. Antenna parameters are summarized in Tab.~\ref{tab:antennas}. Each observing session lasts for about $24$ hours, and sources are observed in blocks of six ten-minute scans throughout the session to better fill the ($u$, $v$) plane (see Fig.~\ref{fig:uv}). Generally, we observe 24 to 30 sources per session. Here we report on the first three epochs of our S band observations (Tab.~\ref{tab:observations}).

Data reduction was carried out in \texttt{AIPS} \citep{greisen03}. Data were loaded using \texttt{FITLD} with \texttt{CLINT} set to $0.1$ minutes and without applying digital corrections. Before fringe fitting, we applied digital sampling corrections, corrected the parallactic angles and calibrated the amplitudes using \texttt{APCAL}. In the case of antennas where system temperatures were not available, we used nominal values to calibrate visibility amplitudes. Delay and rate solutions from \texttt{FRING} were applied using \texttt{CLCAL}. The data was then averaged in frequency, split and written out for imaging in \texttt{Difmap} \citep{shepherd97}. 

After building a source model containing most of the zero-baseline flux, the first amplitude self-calibration was carried out with a solution interval larger than the observing time. After this, cleaning and self-calibration with decreasing solution intervals were iterated until we reached a good dynamic range, and finally, the residuals were added to the final, clean image. First epoch S-band clean images are displayed in Fig.~\ref{fig:cln1}, \ref{fig:cln2} and \ref{fig:cln3}. The first two plots show the TeV-detected sample, while the latter displays calibrators and additional sources included in the observations. We then used the \texttt{modelfit} command in \texttt{Difmap} to model the source structure with circular Gaussian components.

\section{Results}
\label{results}

We have summarized the properties of the twenty-six targets and two calibrators included in the first three S-band observational epochs in Tab.~\ref{tab:source_list}. The table displays the source designation and common name, the source class, and redshift. These observations are focused on the TeV sample, but one neutrino source, $1424-418$ was also added to these sessions. Ten out of twenty-six sources have been studied previously in the X-band by the TANAMI team, so references to these papers are also shown.

Based on the parameters of the \texttt{modelfit} components we can calculate the brightness temperature, $T_{\mathrm{b, obs}}$ of the core components the following way \citep{kovalev05}:
\begin{equation}
    T_{\mathrm{b, obs}} [\mathrm{K}] = 1.22\times 10^{12} \Bigg(\frac{S_{\nu}}{\mathrm{Jy}}\Bigg) \Bigg(\frac{\nu}{\mathrm{GHz}}\Bigg)^{-2} \Bigg(\frac{b_{\mathrm{min}}\times b_{\mathrm{maj}}}{\mathrm{mas}^2}\Bigg)^{-1} (1+z),
\end{equation}
where $S_{\mathrm{\nu}}$ is the flux density of the component, $\nu$ is the observing frequency and $b_{\mathrm{min}}$ and $b_{\mathrm{maj}}$ are the minor and major axes of the component. Errors in flux density are estimated to be $10$\% \citep{ojha10} and errors of the components sizes are taken as one fifth of the beam minor axis \citep{lister09}. We computed the resolution limit based on Eq.~$2$ of \citet{kovalev05}, and we found that all components are resolved (see Tab.~\ref{tab:clns}).

\onecolumn
\begin{figure*}[h!]
    \centering
    \includegraphics[width=0.995\linewidth]{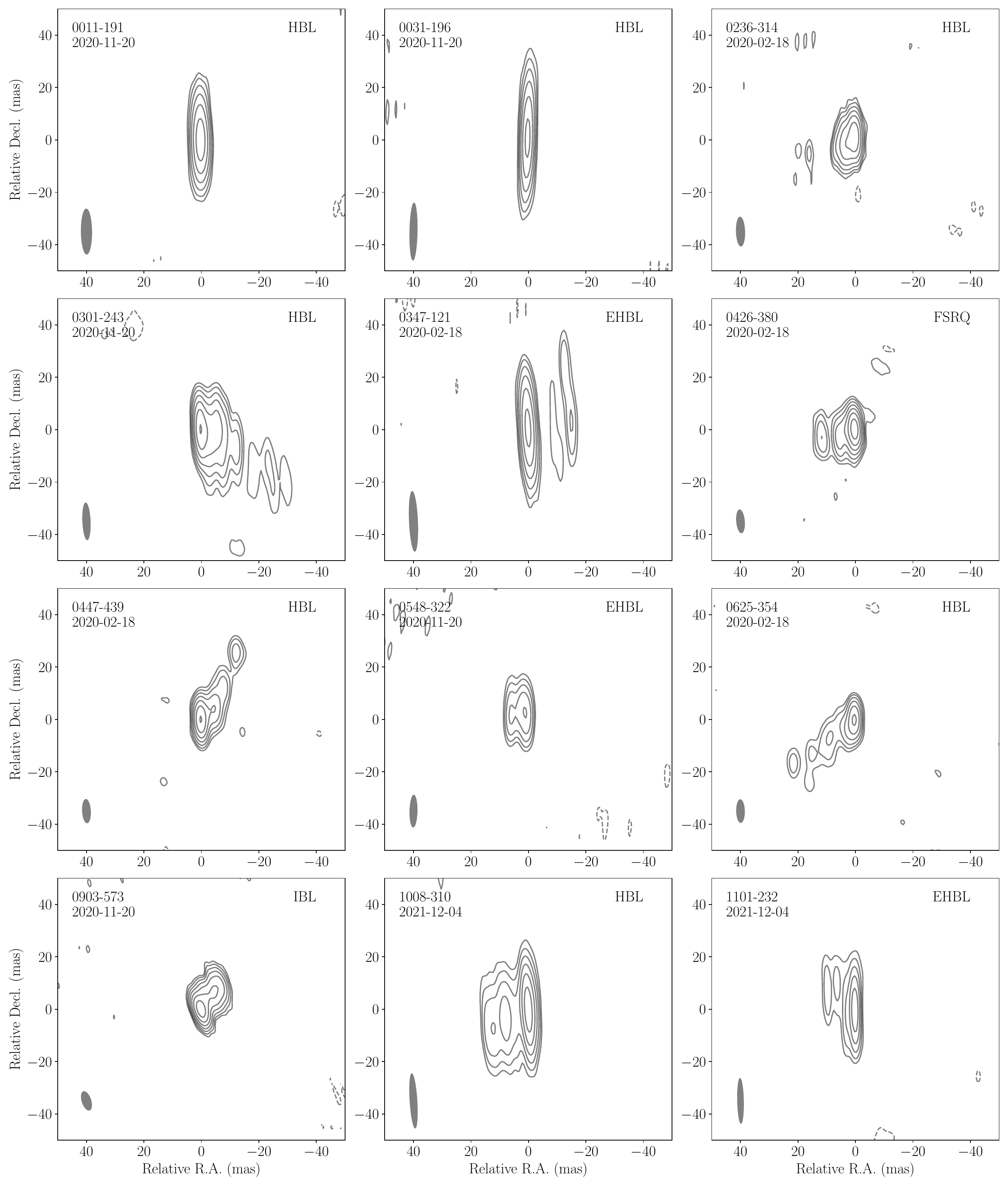}
    \caption{Clean maps of the TANAMI TeV sample at $2.3$\,GHz. Image properties are summarized in Tab.~\ref{tab:clns}. Lowest contours are listed in Tab.~\ref{tab:clns}, and contour levels increase by a factor of two. The class is given at the top right corner of the image (see Tab.~\ref{tab:source_list}).}
    \label{fig:cln1}
\end{figure*}

\begin{figure*}[h!]
    \centering
    \includegraphics[width=0.995\linewidth]{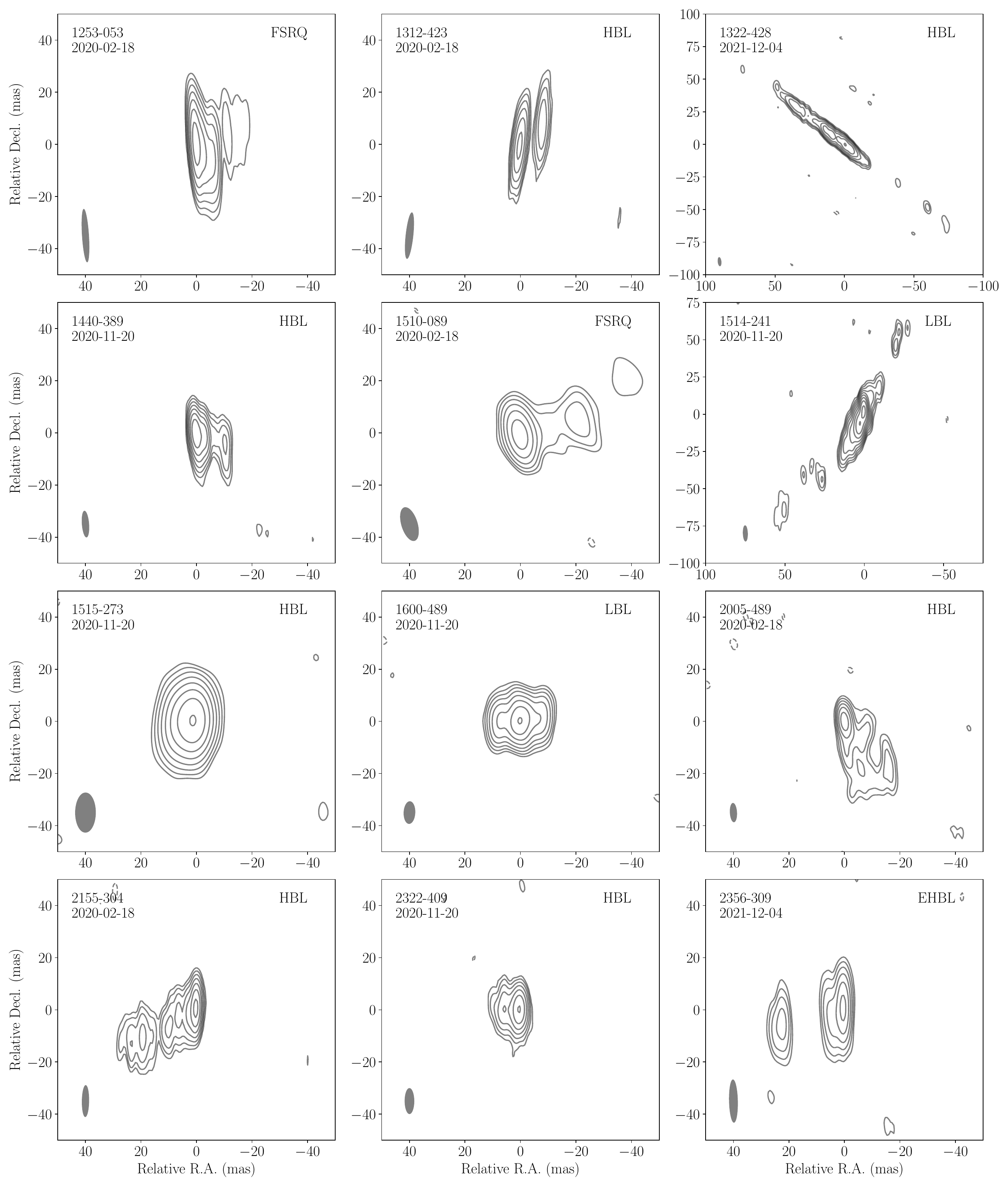}
    \caption{Clean maps of the TANAMI TeV sample at $2.3$\,GHz, continuing Fig.~\ref{fig:cln1}. Lowest contours are listed in Tab.~\ref{tab:clns}, and contour levels increase by a factor of two. The class is given at the top right corner of the image (see Tab.~\ref{tab:source_list}).}
    \label{fig:cln2}
\end{figure*}

\twocolumn
\noindent
Characteristics of the clean hybrid images are summarized in Tab.~\ref{tab:clns} for the TeV sample and in Tab.~\ref{tab:other} for additional sources. The source structure is described in Col.~$2$; Col.~$3$ shows the observing epoch, and Col.~$4-8$ show the clean image properties including beam major and minor axis, position angle, total flux density, the rms noise level and the lowest contours used in the clean images in Fig.~\ref{fig:cln1} and ~\ref{fig:cln2}. Col.~$9-11$ describe the flux density, size and brightness temperature of the core components, and Col.~$12$ and $13$ display the high energy properties of the source.

\section{Notes on individual sources}
\label{sources}

In this section we provide a short summary of previous results on each source in our TeV-sample and describe the source morphology in our observations.

\noindent
\textit{0011$-$191:} $8.4$\,GHz observations of \citet{piner14} with the VLBA reveal a compact core-jet structure with an opening angle of $26.2$\degree. The jet component is close to stationary at $\sim1$~mas \citep{piner18}. Our $2.3$\,GHz images reveal a faint, compact core structure with the flux density of $140$~mJy.

\noindent
\textit{0031$-$196:} 0031$-$196 is part of the HBL sample presented in \citet{piner14}. It shows an extended jet on parsec scales that has a proper motion of $0.204\pm0.071$~mas~yr$^{-1}$ and an opening angle of $18.2$\degree. The source was detected at VHE by H.E.S.S. \citep{becherini12}. Our images show only a core with the flux density of $200$~mJy, because we lack the sensitivity to detect the extended jet emission.

\noindent
\textit{0236$-$314:} 0236$-$314 is an X-ray source from the ROSAT Bright Survey \citep{schwope00}, and has been detected by \textit{Fermi} in 2FGL \citep{ackermann11} and H.E.S.S. \citep{gate17} as well. The TANAMI S band image shows a single-sided core-jet structure.

\noindent
\textit{0301$-$243:} The source has been monitored by MOJAVE at $15$~GHz between $2010$ and $2012$. Superluminal apparent speeds up to $2.3\pm0.5$~$c$ were detected \citep{lister19}. Both the $15$\,GHz MOJAVE and $22$ and $43$\,GHz \citet{whittier} observations show an unresolved core and an extended jet towards the southwest. We recover a similar structure with an extended jet reaching $\sim30$~mas. The source has been part of the \textit{Fermi} Bright Source List \citep{abdo09}, and has been detected by H.E.S.S. \citep{hess0301} with a $9.4\sigma$ significance and a flux of $1.4$\% of that of the Crab Nebula.

\noindent
\textit{0347$-$121:} The source shows a compact structure with a faint jet at $8.4$\,GHz \citep{piner14}, similar to what we see in the S band. The opening angle is measured to be $21.6$\degree, and apparent speeds of $1.7\pm1.2$~$c$ are detected by \citet{piner18}. VHE detection by H.E.S.S. reveals an integral flux of $0.02$~Crab Units (CU) above $250$~GeV and a power--law spectrum with $\Gamma=3.10\pm 0.33$ between $250$~GeV and $3$~TeV \citep{2007A&A...473L..25A}. The SED can be reasonably described by a one-zone model \citep{2007A&A...473L..25A}.

\noindent
\textit{0426$-$380:} 0426$-$380 is currently the furthest known VHE emitting FSRQ at $z=1.1$ \citep{tanaka13}. The source has been part of the TANAMI X-band sample since $2009$, and first images were published in \citet{müller18}. 0426$-$380 shows a core--jet structure with the jet pointing towards the west--southwest.

\noindent
\textit{0447$-$439:} The source is a \textit{Fermi} Bright Source List blazar \citep{abdo09} and have been detected at VHE \citep{hess0447} with the integrated flux density of $0.03$~CU. 0447$-$439 shows a faint extended jet towards the northwest both in S- and X-band \citep{müller18}. 

\noindent
\textit{0548$-$322:} The source resides in a giant elliptical galaxy in the center of Abell S0549 \citep{falomo95}. 0548$-$322 has also been observed by \citet{piner13} and \citet{piner14} at $8$ and $15$\,GHz. These observations reveal an unresolved core and a faint jet pointing towards the northeast with an opening angle of $13.4$\degree. Our $2.3$\,GHz observations, on the other hand, due to the low image sensitivity show only the core with a flux density of $30$~mJy.

\noindent
\textit{0625$-$354:} 0625$-$354 is an FR\,I radio galaxy, hosted by a giant elliptical galaxy in the center of Abell\,$3392$. First TANAMI images of the source at $8.4$\,GHz were published in \citet{ojha10}. Based on the kinematic analysis of nine epochs \citet{angioni19} found superluminal motions with apparent speeds reaching $\sim3$~$c$ from the radio core.
In the S band the jet extends $\sim20$~mas to the southeast, but tapered images of \citet{ojha10} reveal a jet up to $40$~mas. Among many other FR\,I radio galaxies, 0625-354 has also been detected at very high energies \citep{dyrda15} with an integral flux of $0.04$~CU and photon index of $\Gamma=2.8\pm0.5$.

\noindent
\textit{0903$-$573:} 0903$-$573 was first detected in the GeV range by \textit{Fermi} LAT in $2015$ \citep{ojha15}. The source has been considered a blazar candidate of unknown type until recently, when it was classified as a BL\,Lac object by \citet{pita17}. 0903$-$573 exhibited a complex $\gamma$-ray flaring activity during $2018$ and $2020$ \citep{mondal21} that coincides with our first S-band observation. Our image reveals a core and an extended jet in the northwest direction.

\noindent
\textit{1008$-$310:} X band observations of \citet{piner14} and \citet{piner18} show a compact core structure with a faint jet with opening angle of $47.8$\degree. Our $2.3$\,GHz image shows a slightly more extended jet towards the east--northeast. Integral flux above $0.2$~TeV is $0.008$~CU \citep{hess1008}.

\noindent
\textit{1101$-$232:} The source is hosted by an elliptical host galaxy. On kiloparsec scales it shows a one-sided structure with a diffuse extension to the north \citep{laurent93}. $5$ and $8$\,GHz observations with the VLBA \citep{tiet12} as well as our $2.3$\,GHz TANAMI observations show a core with a faint, slightly extended jet to the northwest. \citet{aharonian07} has detected the source at VHE with H.E.S.S. SED modeling \citep{yan12} can recover the observed TeV emission by using slightly beamed high-energy jets emission, but only the models with moderately high Doppler factors result in a jet in equipartition. \citet{yan12} suggests that inverse Compton scattering of the CMB photons in the extended jet is the main source of the TeV emission.

\noindent
\textit{1253$-$053:} 3C\,$279$ is one of the most well-studied AGN in VLBI science. It has been monitored by MOJAVE at $15$\,GHz since $1995$, and by the Boston University Blazar Group at $43$\,GHz since $2007$. On VLBI scales the jet has a well-collimated one-sided structure -- similar to our S-band images--, however on kiloparsec scales it shows a significant bend towards the east \citep{cheung02}. \textit{RadioAstron} observations at $22$\,GHz reveal a filamentary structure in the jet interpreted as Kelvin--Helmholtz instability \citep{fuentes22}. Event Horizon Telescope (EHT) observations at $230$\,GHz show two components within the inner $150$~$\mu$as of the $86$\,GHz core, with the upstream feature identified as the core oriented perpendicular to the downstream jet component \citep{eht279}. The core features show apparent speeds of $\sim15-20$~$c$, which are consistent with values measured at cm wavelengths. 3C\,$279$ was the first blazar detected in the $\gamma$-rays \citep{hartman92}, and has since been detected at VHE by MAGIC as well \citep{magic08}. The source has also been noted as a possible source of astrophysical neutrinos \citep{plavin20}.

\noindent
\textit{1312$-$423:} 1312$-$423 has been detected by H.E.S.S. \citep{hess1312} with the integral flux density of $0.5$\% of that of the Crab Nebula, and with the careful modelling of Cen\,A they recover the source in the high-energy \textit{Fermi} LAT data as well. Our S-band observations show a core and a slight extension of the jet towards the northeast.

\noindent
\textit{1322$-$428:} Cen\,A is an FR\,I radio galaxy \citep{1998AJ....115..960T}, and is the closest radio-loud AGN to the Earth. The source was detected early on by \textit{Fermi} LAT and is part of the \textit{Fermi} Bright Source List \citep{abdo09}. The first X-band TANAMI image of Cen\,A was published in \citet{ojha10}, and it was followed by a detailed analysis on the source by \citet{müller14}. The jet shows differential motion, that is to say, downstream features exhibit higher velocities than the ones closer to the VLBI core. This observation lead the authors to propose a spine-sheath structure present in the jet, which have been detected on sub-parsec scales at $230$\,GHz with the EHT \citep{janssen21}. Based on the jet-counterjet ratio the viewing angle is constrained to fall between $12$\degree and $45$\degree. Our S-band images also show a double-sided structure with an extended, well-collimated jet.

\noindent
\textit{1440$-$389:} $8.4$\,GHz images from TANAMI were presented in \citet{müller18}. The source shows a compact structure in S band with a jet that can be modeled by a single component. 1440$-$389 has first been detected at VHE in 2012 \citep{hofmann12}, and have been recently detected in a high $\gamma$-state by the \textit{Fermi} LAT for the first time \citep{ciprini22}.

\begin{figure*}[h!]
    \centering
    \includegraphics[width=\linewidth]{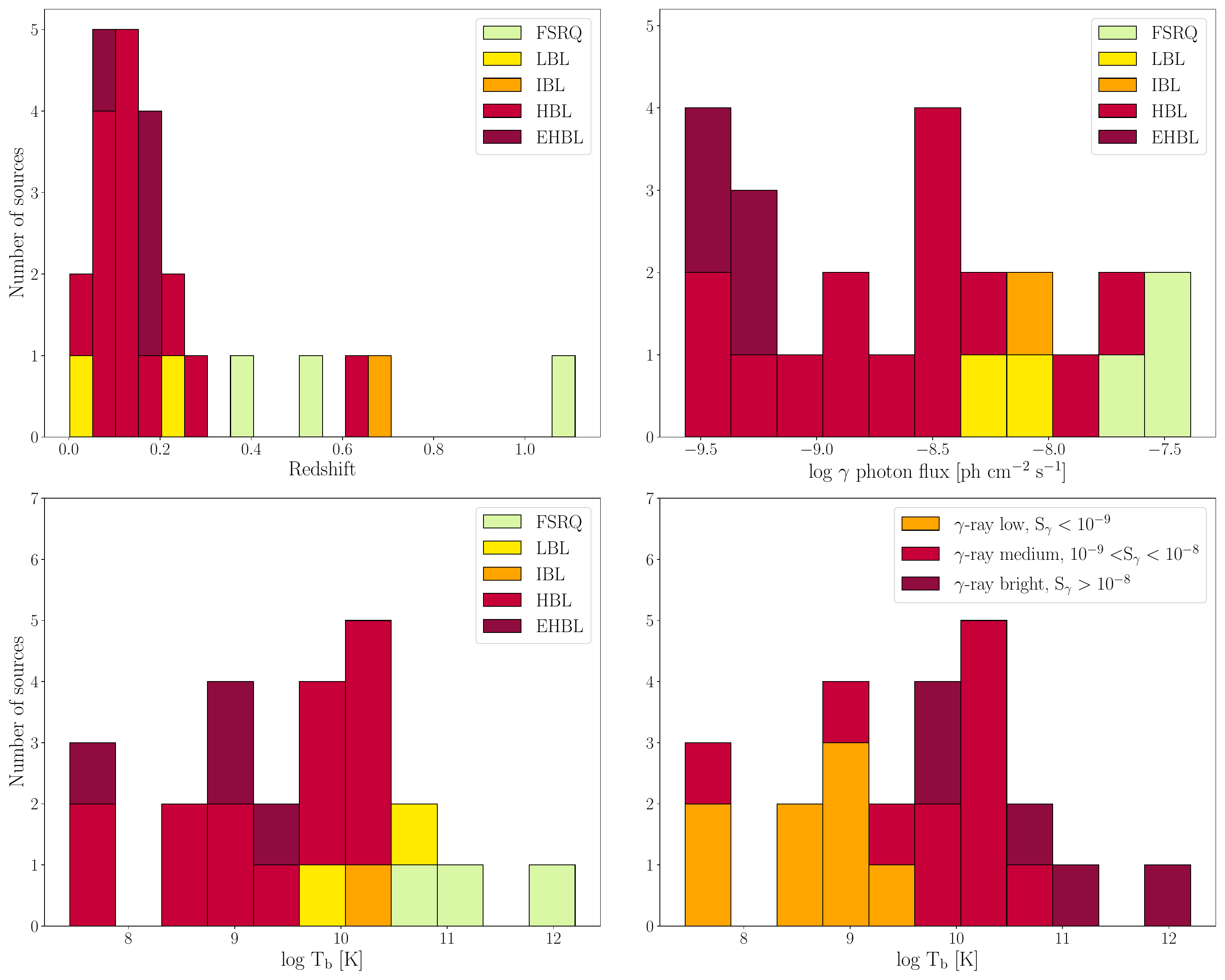}
    \caption{Histogram showing the redshift (upper left panel), $0.1-100$~GeV photon flux (upper right panel) and $2.3$-GHz core brightness temperature (lower panels) distribution of sources in the TeV sample.}
    \label{fig:hist}
\end{figure*}

\noindent
\textit{1510$-$089:} 1510$-$089 has been part of the MOJAVE program since $1995$, but observations were dropped in $2013$, and have been re-taken recently in $2021$. On parsec scales the jet shows a one-sided structure, which we also recover in our $2.3$-GHz images. The MOJAVE kinematic analysis \citep{lister19} shows maximum apparent speeds up to $28.0\pm0.6$~$c$, which suggests a highly beamed emission \citep{liodakis17}. Multi-frequency radio observations during high states in $2015-2017$ reveal the emergence of two new jet components nearly simultaneously to the high-energy flares, and that during one of the flares the jet showed a limb-brightened linear polarization structure in the core region \citep{2019ApJ...877..106P}. The authors conclude that the passing of the new components through stationary shocks gives rise to the $\gamma$-ray emission, and that the jet shows a complex transverse velocity structure that changes with time \citep{2015ApJ...804..111M, 2017Galax...5...67C}.

\noindent
\textit{1514$-$241:} AP\,Librae has been part of MOJAVE between $1997$ and $2012$. $15$\,GHz observations show a $30$~mas extended jet with a complex structure. In S band we recover a double-sided jet. The maximum jet speed from MOJAVE observations is $6.21\pm0.26$~$c$ \citep{lister19}. If we assume Doppler factors above $10$, this suggests the jet viewing angle to fall below $5$\degree \citep{2013AJ....146..120L}. Given the superluminal apparent jet speed and the small viewing angle, the double-sided jet structure in the S band image is of interest. VHE detection was reported by \citet{hofmann10} with H.E.S.S. The VHE emission is proposed to originate from the extended jet --that is also detected in the X-rays-- via IC scattering of the CMB photons \citep{zacharias16}.

\noindent
\textit{1515$-$273:} TXS\,1515$-$273 was detected by MAGIC \citep{mirzoyan19} during a $\gamma$-ray flare by \textit{Fermi} \citep{cutini19}. MOJAVE images reveal a one-sided jet structure that can be modeled by one component with an apparent speed of $0.73\pm0.18$~$c$ \citep{lister21}. Our S-band images also show a core-jet morphology pointing towards the east.

\noindent
\textit{1600$-$489:} First $8.4$-GHz TANAMI observations of the object were published in \citet{müller14}. The source shows a double-sided structure and due to its non-blazar like spectral properties it is classified as a compact symmetric object (CSO). The source is bright in the $\gamma$-rays \citep{abdo09}. Our new S-band images showing a double-sided jet structure and bright central core are also reminiscent of the CSO class.

\noindent
\textit{2005$-$489:} The source was the first blazar independently detected by H.E.S.S. at very high energies, and the second source noted as a VHE emitter in the southern hemisphere \citep{2005A&A...436L..17A}. The integral flux is $0.025$ of that of the Crab Nebula and has a very soft spectrum with photon index of $\Gamma=4.0\pm0.4$. X-band TANAMI observations of \citet{ojha10}, as well as our new $2.3$\,GHz image shows an elongated core in the northeast-southwest direction and a diffuse jet emission to the southwest.

\noindent
\textit{2155$-$304:} $15$\,GHz VLBA \citep{piner04,piner10} , $8$\,GHz and $22$\,GHz TANAMI observations \citep{ojha10}, as well as the new $2.3$\,GHz image reveal an initially well-collimated jet on mas scales. The first polarimetric image published by \citep{piner08} reveals polarization in the core region with polarization fraction of $2.9$\% and EVPAs with position angle of $131$\degree, misaligned with the jet position angle by $30$\degree. 2155$-$304 has been detected at VHE already in $1999$ by the University of Durham Mark 6 atmospheric Cherenkov telescope \citep{chadwick99}, and can exhibit integral fluxes up to $60$\% of that of the Crab Nebula \citep{aharonian05}. The source exhibited a high flaring activity during $2006$, when Doppler factors over $100$ were used to model the SED \citep{2007ApJ...664L..71A}, which is much higher than typical Doppler factors estimated from VLBI data. 

\noindent
\textit{2322$-$409:} While observing PKS\,2316$-$423, \citet{abdalla19} detected an excess in hard $\gamma$-rays in the position of 1ES\,2322$-$409 with H.E.S.S. The VHE spectrum can be described with a power-law with the soft photon index of $\Gamma=3.40\pm0.86$. The only published VLBI image of the source is presented by \citet{schinzel17} with the VLBA at $7.62$\,GHz. The source has a jet extending $\sim4$~mas towards the northeast. Our S band image shows a similar morphology, but the jet points towards east.

\noindent
\textit{2356$-$309:} The source has been reported as an extreme synchrotron peaked blazar since its low-energy SED component peaks in the X-rays \citep{costamante01}. VLBI observations of \citet{tiet12} at $5$ and $8$\,GHz show a core and a jet extending a few mas from the core. Our new TANAMI images reveal a similar structure.

\section{Discussion}
\label{disc}

The morphology of radio-loud AGN can be described as double-sided, when we detect both the jet and the counter-jet, single-sided, when we only see the jet, and as compact cores without jet emission \citep{1998AJ....115.1295K}. $25$\% of the sources in the sample show only the radio core, $62.5$\% are single-sided, and the rest only exhibit double-sided jets. Faint $S_{\mathrm{2.3\,GHz}}<50$~mJy sources on average show no extended jet emission, because the low SNR prohibits the robust detection of jet emission. With that said, $9$ targets have total flux densities below $50$~mJy, and only $5$ reach $1$~Jy.
 
The upper left-hand panel of Fig.~\ref{fig:hist} shows the redshift distribution of the TANAMI TeV sources. While the distribution of HBL and EHBL objects peak around $z\sim0.1$, FSRQ in the sample fall between $z\sim0.36$ and $1.11$, and 0426$-$380 at $z=1.11$ is the furthest known TeV emitter to date \citep{tanaka13}. The distribution of BL\,Lacs and FSRQ is consistent with the redshift distribution seen in the Bright AGN Source List \citep{abdo09b} -- which is a $\gamma$-ray selected sample--, where BL\,Lacs are mainly located below redshift $0.5$, and the distribution of FSRQ peak around $z\sim1$. However, the redshift distribution of TeV-detected AGN is biased due to several reasons. The current generation of Imaging Atmospheric Cherenkov Telescopes (IACTs) operate in pointing mode, which means that they lack full-sky coverage. In addition, most AGN observed at VHE are only detected during high states of activity due to the lack of sensitivity of IACTs. These issues, however, will be overcome with the arrival of the Cherenkov Telescope Array \citep{2011ExA....32..193A}. TeV emission from high-redshift sources is also affected by the attenuation from the extragalactic background light. In addition, many BL\,Lac objects have no redshifts available due to the lack of detectable emission lines in their spectra, especially when it comes to high-redshift sources, further decreasing the number of known distant TeV-emitter AGN \citep{2018JApA...39...52S}.

The logarithmic GeV $\gamma$-ray photon flux distribution (Fig.~\ref{fig:hist}, upper right panel) show that BL\,Lac objects are distributed evenly, but mainly inhabit the fainter end of the sample. In addition, $\gamma$-ray brightness shows an inverse relation with the frequency of the synchrotron peak, that is, EHBL and HBL sources are dimmer at high energies. The photon flux distribution of the \textit{Fermi}-detected blazars from the above mentioned Bright AGN Source List \citep{abdo09b} peaks around $10^{-7}$~ph cm$^{-2}$ s$^{-1}$. However, our TeV sample shows a rather uniform distribution with the mean photon flux of $9.09\times10^{-9}$~ph cm$^{-2}$ s$^{-1}$.

The 2.3 GHz core brightness temperature distributions (Fig.~\ref{fig:hist}, lower panels) show that BL\,Lac objects have the lowest T$_{\mathrm{b, core}}$ values, with the mean of $7.49\times10^9$~K, and FSRQ exhibit the highest T$_{\mathrm{b, core}}$ values, with the average of $1.52\times10^{10}$~K. This is also seen in the study of a complete flux-limited sample of AGN \citep{2011ApJ...742...27L}, where HBL sources show lower brightness temperatures then other BL\,Lac objects and quasars. \citep{homan21} reports that BL\,Lac objects in this flux-limited sample show the same trend that we found for our TeV-detected sample, namely that there is an inverse relation between the synchrotron peak frequency and the median core brightness temperature.

T$_{\mathrm{b, core}}$ distribution with $\gamma$-ray flux classes marked shows that $\gamma$-ray low (S$_{\mathrm{\gamma}}<10^{-9}$~ph cm$^{-2}$ s$^{-1}$) objects tend to exhibit lower core brightness temperatures, while medium ($10^{-9}<$S$_{\mathrm{\gamma}}<10^{-8}$~ph cm$^{-2}$ s$^{-1}$) and bright ($>10^{-8}$~ph cm$^{-2}$ s$^{-1}$) sources have higher T$_{\mathrm{b, core}}$ values. A similar trend is recovered for the above-mentioned MOJAVE $1.5$-Jy flux-limited sample, in which quasars show higher median core brightness temperatures and are more luminous at GeV energies than BL\,Lacs \citep{homan21}. These results suggest that the emission from FSRQ is more strongly Doppler boosted than from BL\,Lacs, especially from HBL sources.

The average and median $T_{\mathrm{b, core}}$ for the TANAMI TeV sample are $8.19\times10^{10}$~K and $4.9\times10^{9}$~K, respectively. Twenty-one $T_{\mathrm{b, core}}$ values are below the equipartition brightness temperature of $T_{\mathrm{eq}}\approx5\times10^{10}$~K \citep{readhead94} and all except one value falls below the inverse Compton catastrophe limit of $\approx10^{12}$~K \citep{keller69}. This suggests that in most cases the emission from these sources is not beamed at $2.3$\,GHz.
$15$\,GHz observations of MOJAVE, a sample dominated by LBL and FSRQ objects, reveal a mean core brightness temperature of $2.93\times10^{11}$~K with a median of $3.66\times10^{11}$~K \citep{homan21}, which is significantly higher than the ones observed in our sample. 
$T_{\mathrm{b, core}}$ values of the X-band TANAMI sample from \citet{bock16} have an average and median of $9.62\times10^{11}$~K and $2.1\times10^{11}$~K, which are consistent with those of the MOJAVE sample. The nearly one order of magnitude lower average $T_{\mathrm{b, core}}$ for the TeV sample has been shown as a characteristic of VHE detected sources \citep{2018mgm..conf.3074P}.

\section{Conclusion}
\label{summ}

In this work we present first results of our new $2.3$\,GHz AGN monitoring including all $24$ TeV-detected sources at the southern hemisphere at the time of the observations. The $2.3$\,GHz band was added in our program to better study extended jet emission and to include TeV sources in our sample that would be too faint to detect at higher frequencies. With the inclusion of Parkes and ATCA, we were capable of detecting even the faintest ($S_{\mathrm{2.3G\,Hz}}\approx10$~mJy) target in the sample. While this was a success, the current long transoceanic baselines to the Hartebeesthoek antenna yield only limited sensitivity, which degrades the image fidelity and angular resolution in observations of faint targets. This is a strong limitation for studies of the bulk of the HBL and extreme-blazar population. In order to improve our observations, we support the efforts to phase up the MeerKAT array, with which we could achieve three times better sensitivity on the long baselines.

We have fitted circular Gaussian components to model the source structures and calculated the brightness temperatures for the radio core. Core brightness temperature distributions reveal that FSRQ in the sample generally have higher core brightness temperatures then BL\,Lacs and radio galaxies. We also observe that with increasing $\gamma$-ray photon flux core brightness temperatures are higher as well. Comparing this sample to $8.4$\,GHz TANAMI and $15$\,GHz MOJAVE results, which samples are dominated by FSRQ and low synchrotron peaked objects, we find that the $T_{\mathrm{b, core}}$ values in the S band are up to a magnitude lower than average values in the other samples, which is consistent with previous results of \citet{2018mgm..conf.3074P} stating that TeV-detected sources have lower $T_{\mathrm{b, core}}$. However, at low frequencies it is also possible that we do not detect the optically thick VLBI core, only a bright, optically thin jet component that we identify as the core. Since brightness temperature measurements are heavily dependent on the maximum baseline length and the flux density of the component, bright jet features, like the core, can exhibit high $T_{\mathrm{b}}$ values.

Due to the success of the first three S band observing epochs, we continue to monitor VHE-detected AGN and perform kinematic and spectral studies once a sufficient number of observing epochs is reached. While the $2.3$\,GHz observations lack the resolution and sensitivity to determine the location of the $\gamma$-ray production or to differentiate between jet models suggested to resolve the Doppler crisis, they enable us to study the faintest TeV-emitters and to perform multi-frequency studies together with our $8.4$\,GHz monitoring. TANAMI is currently the only AGN monitoring program on the southern hemisphere, and forming complete samples to study out of these rarely observed targets is crucial to advance multi-wavelength studies.

\vspace{1cm}
\begin{acknowledgements}
The authors would like to thank the anonymous referee for their valuable comments on the manuscript. We thank H. Müller for his constructive suggestions to improve our work.
The Long Baseline Array is part of the Australia Telescope National Facility (\url{https://ror.org/05qajvd42}) which is funded by the Australian Government for operation as a National Facility managed by CSIRO. From the 2023 July 1 operation of Warkworth was transferred from AUT
University to Space Operations New Zealand Ltd with new funding from Land Information New Zealand (LINZ).
This research was supported through a PhD grant from the International Max Planck Research School (IMPRS) for Astronomy and Astrophysics at the Universities of Bonn and Cologne. M2FINDERS project has received funding from the European Research Council (ERC) under the European Union’s Horizon 2020 research and innovation programme (grant agreement No.\,101018682). M.K. and F.R. acknowledge funding by the Deutsche Forschungsgemeinschaft (DFG, German Research Foundation) - grant 434448349.
\end{acknowledgements}

\newpage
\begin{appendix}
\onecolumn

\section{Additional data}

\begin{table*}[h]
    \centering
    \caption{Sources observed during the first three S band epochs.}
    \label{tab:source_list}
    \begin{tabular}{l l c c c c}
    \hline\hline
       Name  & Alternative name & Class\tablefootmark{a} & $z$ & X band first epoch & Association  \\
       \hline
   0011$-$191 & SHBL\,J001355.9$-$185406 & HBL & $0.095$ (1) & - & TeV \\
   0031$-$196 & KUV\,00311$-$1938 & HBL & $0.610$ (2) & - & TeV \\
   0236$-$314 & 1RXS\,J023833.1$-$311808 & HBL & $0.233$ (3) & - & TeV \\
   0301$-$243 & & HBL & $0.260$ (2)& - & TeV \\
   0347$-$121 & 1ES\,0347$-$121 & EHBL & $0.185$ (2) & - & TeV \\
   0426$-$380 & & FSRQ & $1.111$ (2) & (16) & TeV \\
   0447$-$439 & & HBL & $0.107$ (4) & (16) & TeV \\
   0548$-$322 & & EHBL & $0.069$ (2) & - & TeV \\
   0625$-$354 & OH$-$342 & RG & $0.056$ (5) & (17) & TeV \\
   0903$-$573 & & IBL & $0.695$ (6) & - & TeV \\
   1008$-$310 & 1RXS\,J101015.9$-$311909 & HBL & $0.143$ (1) & - & TeV \\
   1101$-$232 & 1ES\,1101$-$232 & EHBL & $0.186$ (7) &- & TeV \\
   1248$-$350 & & U & $0.410$ (8) & - & neutrino \\
   1253$-$053 & 3C\,279 & FSRQ & $0.538$ (7) & - & neutrino/TeV \\
   1312$-$423 & 1ES\,1312$-$423 & HBL & $0.105$ (2) & - & TeV \\
   1322$-$428 & Cen\,A, NGC\,5128 & RG & $0.002$ (9) & (17) & TeV\\
   1424$-$418 & [HB89]\,1424$-$418 & FSRQ & $1.520$ (10) & (17) & neutrino \\
   1440$-$389 & & HBL & $0.065$ (11) & (16) & TeV \\
   1510$-$089 & & FSRQ & $0.360$ (10) & - & TeV \\
   1514$-$241 & AP\,Librae & LBL & $0.049$ (1) & - & TeV \\
   1515$-$273 & TXS\,1515$-$273 & HBL & $0.140$ (11) & - & TeV \\
   1600$-$489 & PMN\,J1603$-$4904 & CSO & $0.232$ (12) & (18) & TeV \\
   1921$-$293 & & BL\,Lac & $0.353$ (1) & - & calibrator \\
   1934$-$638 & & CSO & $0.183$ (13) & (19) & calibrator \\
   2005$-$489 & [HB89]\,2005$-$489 & HBL & $0.071$ (14) & (17) & TeV\\
   2155$-$304 & [HB89]\,2155$-$304 & HBL & $0.116$ (15) & (17) & TeV \\
   2322$-$409 & 1ES\,2322$-$409 & HBL & $0.174$ (1) & - & TeV \\
   2356$-$309 & & EHBL & $0.165$ (1) & - & TeV \\
      \hline
    \end{tabular}
    \tablefoot{
    \tablefoottext{a}{Source class abbreviations: LBL low frequency peaked blazar; IBL intermediate frequency peaked blazar; HBL high frequency peaked blazar; EHBL extremely high frequency peaked blazar; RG - radio galaxy; FSRQ - flat spectrum radio quasar; CSO - compact symmetric object; U - unidentified.}
    \tablebib{(1) \citet{2009MNRAS.399..683J}; (2) \citet{2011NewA...16..503M}; (3) \citet{2015A&A...579A..34A}; (4) \citep{1997AJ....114.1356C}; (5) \citep{2014ApJ...797...82L}; (6) \citep{1990PASP..102.1235T}; (7) \citep{2016ApJ...818..113N}; (8) \citep{2002A&A...386...97J}; (9) \citep{2011MNRAS.416.2840L}; (10) \citep{2017ApJS..233....3T}; (11) \citep{2019A&A...632A..77C}; (12) \citep{2016A&A...586L...2G}; (13) \citep{2002LEDA.........0P}; (14) \citep{2018ApJS..237...11K}; (15) \citep{2013MNRAS.435.1233G}; (16) \citet{müller18}; (17) \citet{ojha10}; (18) \citet{müller14}; (19) \citet{ojha04}.}
    }

\end{table*}

\newpage

\begin{figure*}[h!]
    \centering
    \includegraphics[width=0.7\linewidth]{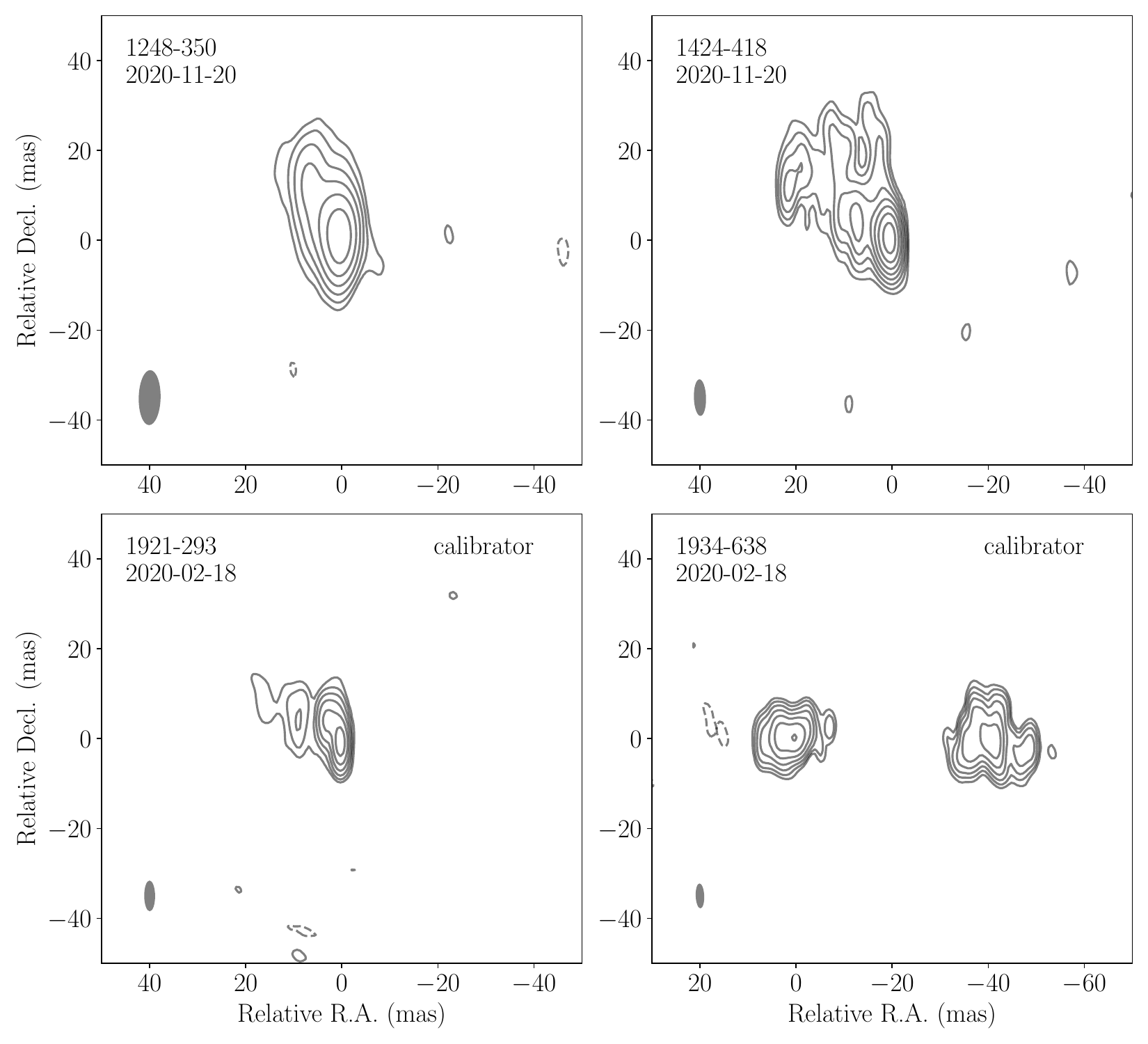}
    \caption{Clean maps of additional sources and calibrators included in the $2.3$\,GHz observations. Image parameters and lowest contours are listed in Tab.~\ref{tab:other}. Contour levels increase by a factor of two.}
    \label{fig:cln3}
\end{figure*}

\begin{table*}[h]
    \centering
    \caption{Image properties of non-TeV sources and calibrators observed during 2020-2021 in S band.}
    \label{tab:other}
    \begin{tabular}{l c l c c c c c c}
    \hline\hline
    Source & Morphology\tablefootmark{a} & Epoch & b$_{\mathrm{maj}}$\tablefootmark{b} &  b$_{\mathrm{min}}$\tablefootmark{c} & PA\tablefootmark{d} & S$_{\mathrm{tot}}$\tablefootmark{e} & $\sigma$\tablefootmark{f} & $I_{\mathrm{base}}$\tablefootmark{g} \\
     & & [yyyy-mm-dd] & [mas] & [mas] & [\degree] & [mJy] & [mJy/beam] & [mJy/beam]\\
    \hline
        1248$-$350 & SS & 2020-11-20 & 12.2 & 4.6 & $-1$ & 200.8 & 0.6 & 2.2\\
        1424$-$418 & SS & 2020-11-20 & 8.1 & 2.5 & 2 & 1988.0 & 1.6 & 5.1 \\
        1921$-$293 & SS & 2020-02-18 & 6.84 & 2.3 & 0 & 5087.7 & 14.9 & 47.3 \\
        1934$-$638 & DS & 2020-02-18 & 5.54 & 1.8 & 2 & 11194.5 & 5.2 & 17.9 \\
    \hline
    \end{tabular}
    \tablefoot{
    \tablefoottext{a}{Source morphology: SS - single-sided; DS - double-sided jet morphology.}
    \tablefoottext{b}{Beam major axis.}
    \tablefoottext{c}{Beam minor axis.}
    \tablefoottext{d}{Beam position angle.}
    \tablefoottext{e}{Total flux density.}
    \tablefoottext{f}{Rms noise of the image.}
    \tablefoottext{g}{Lowest contour.}
    }
\end{table*}

\newpage

\begin{sidewaystable*}[h!]
    \centering
    \caption{Image properties of the TeV-detected AGN in our observations.}
    \label{tab:clns}
    \resizebox{\textwidth}{!}{%
    \begin{tabular}{l c l c c c c c c c c c c c}
    \hline\hline
    \small
     Source & Morphology\tablefootmark{a} & Epoch & b$_{\mathrm{maj}}$\tablefootmark{b} &  b$_{\mathrm{min}}$\tablefootmark{c} & PA\tablefootmark{d} & S$_{\mathrm{tot}}$\tablefootmark{e} & $\sigma$\tablefootmark{f} & $I_{\mathrm{base}}$\tablefootmark{g} & S$_{\mathrm{core}}$\tablefootmark{h} & $\Theta_{\mathrm{core}}$\tablefootmark{i} & T$_{\mathrm{b}}$\tablefootmark{j} & S$_{\mathrm{0.1-100 GeV}}$\tablefootmark{k} & S$_{\mathrm{TeV}}$\tablefootmark{l}\\
      & & [yyyy-mm-dd] & [mas] & [mas] & [\degree] & [mJy] & [mJy/beam] & [mJy/beam] & [mJy] & [mas] & [K] & [ph/cm$^2$/s] & CU\\
     \hline
   0011$-$191 & C & 2020-11-20 & $17.6$ & $3.9$ & $1$ & $16.4$ & $14.4$ & $0.3$ & $16.6\pm1.7$ & $8.8\pm0.8$ & $(5.5\pm1.1)\times10^{7}$ & $(2.71\pm0.33)\times10^{-10}$ & $0.006$\\
   0031$-$196 & C & 2020-11-20 & $22.2$ & $2.8$ & $-1$ & $19.1$ & $14.5$ & $0.4$ & $19.1\pm1.9$ & $14.2\pm0.6$ & $(3.6\pm0.5)\times10^{7}$ & $(2.69\pm0.08)\times10^{-9}$ & $0.009$ \\
   0236$-$314 & SS & 2020-02-18 & $11.4$ & $3.2$ & $2$ & $24.9$ & $14.8$ & $0.3$ & $14.9\pm1.5$ & $3.3\pm0.6$ & $(4.0\pm1.6)\times10^{8}$ & $(9.13\pm0.48)\times10^{-10}$ & -\\
   0301$-$243 & SS & 2020-11-20 & $14.4$ & $2.9$ & $2$ & $170.5$ & $72.2$ & $1.1$ & $91.9\pm9.2$ & $1.4\pm0.6$ & $(1.4\pm1.2)\times10^{10}$ & $(4.79\pm0.11)\times10^{-9}$ & $0.014$ \\
   0347$-$121 & SS & 2020-02-18 & $23.3$ & $3.1$ & $2$ & $10.4$ & $7.1$ & $0.2$ & $10.0\pm1.0$ & $10.0\pm0.6$ & $(2.8\pm0.4)\times10^{7}$ & $(3.69\pm0.37)\times10^{-10}$ & $0.02$\\
   0426$-$380 & SS & 2020-02-18 & $9.1$ & $3$ & $4$ & $761.1$ & $527.1$ & $5.3$ & $599.1\pm59.9$ & $1.5\pm0.6$ & $(1.4\pm1.1)\times10^{11}$ & $(2.14\pm0.02)\times10^{-8}$ & -\\
   0447$-$439 & SS & 2020-02-18 & $9.3$ & $3.1$ & $3$ & $89.9$ & $53.6$ & $0.8$ & $58.2\pm5.8$ & $1.9\pm0.6$ & $(4.3\pm2.9)\times10^{9}$ & $(1.25\pm0.02)\times10^{-8}$ & $0.03$\\
   0548$-$322 & C & 2020-11-20 & $10.6$ & $3.0$ & $1$ & $27.17$ & $11.08$ & $0.66$ & $13.94\pm1.39$ & $1.80\pm0.55$ & $(1.1\pm0.7)\times10^{9}$ & $(3.13\pm0.34)\times10^{-10}$ & $0.013$\\
   0625$-$354 & SS & 2020-02-18 & $9.0$ & $3.0$ & $1$ & $352.0$ & $191.9$ & $5.3$ & $218.0\pm21.8$ & $1.8\pm0.6$ & $(1.7\pm1.2)\times10^{10}$ & $1.26\pm0.06)\times10^{-9}$ & $0.04$ \\
   0903$-$573 & SS & 2020-11-20 & $7.6$ & $3.4$ & $16$ & $346.8$ & $181.9$ & $1.8$ & $196.8\pm19.7$ & $2.4\pm0.7$ & $(1.4\pm0.8)\times10^{10}$ & $(9.09\pm0.15)\times10^{-9}$ & - \\
   1008$-$310 & C & 2021-12-04 & $21.2$ & $2.7$ & $3$ & $28.7$ & $14.5$ & $0.3$ & $19.0\pm1.9$ & $2.3\pm0.5$ & $(9.0\pm4.0)\times10^{8}$ & $(6.08\pm0.54)\times10^{-10}$ & $0.008$\\
   1101$-$232 & C & 2021-12-04 & $17.6$ & $2.3$ & $1$ & $38.3$ & $23.6$ & $0.9$ & $30.3\pm3.0$ & $2.2\pm0.5$ & $(1.7\pm0.7)\times10^{9}$ & $(4.29\pm0.44)\times10^{-10}$ & 0.02\\
   1253$-$053 & SS & 2020-02-18 & $20.6$ & $2.5$ & $3$ & $8133.5$ & $4508.6$ & $45.1$ & $5220.1\pm522.0$ & $1.1\pm0.5$ & $(1.6\pm1.4)\times10^{12}$ & $(4.09\pm0.03)\times10^{-8}$ & -\\
   1312$-$423 & C & 2020-02-18 & $18.0$ & $2.8$ & $-6$ & $11.9$ & $7.8$ & $0.2$ & $8.3\pm0.8$ & $2.0\pm0.6$ & $(5.1\pm2.8)\times10^{8}$ & $(3.85\pm0.47)\times10^{-10}$ & 0.005\\
   1322$-$428 & DS & 2021-12-04 & $7.1$ & $2.7$ & $5$ & $3630.6$ & $431.2$ & $8.6$ & $519.8\pm52.0$ & $2.3\pm0.5$ & $(2.3\pm1.1)\times10^{10}$ & $(3.59\pm0.09)\times10^{-9}$ & 0.08\\
   1440$-$389 & SS & 2020-11-20 & $10.3$ & $2.6$ & $4$ & $133.5$ & $78.2$ & $0.6$ & $104.7\pm10.5$ & $1.7\pm0.5$ & $(9.0\pm5.0)\times10^{9}$ & $(4.08\pm0.11)\times10^{-9}$ & 0.054\\
   1510$-$089 & SS & 2020-02-18 & $13.4$ & $5.9$ & $16$ & $2445.4$ & $1524.9$ & $30.5$ & $1842.5\pm184.3$ & $3.6\pm1.2$ & $(4.5\pm2.9)\times10^{10}$ & $(3.38\pm0.03)\times10^{-8}$ & 0.03\\
   1514$-$241 & DS & 2020-11-20 & $11.0$ & $3.1$ & $2$ & $2571.3$ & $960.3$ & $2.9$ & $993.7\pm99.4$ & $2.2\pm0.6$ & $(5.0\pm2.9)\times10^{10}$ & $(6.67\pm0.13)\times10^{-9}$ & 0.04\\
   1515$-$273 & SS & 2020-11-20 & $15.4$ & $7.5$ & $0$ & $120.6$ & $86.9$ & $0.7$ & $81.0\pm8.1$ & $3.7\pm1.5$ & $(1.5\pm1.2)\times10^{9}$ & $(1.71\pm0.07)\times10^{-9}$ & -\\
   1600$-$489 & DS & 2020-11-20 & $8.7$ & $4.2$ & $-2$ & $1282.9$ & $507.6$ & $3.8$ & $581.4\pm58.1$ & $5.4\pm0.8$ & $(5.7\pm1.9)\times10^{9}$ & $(5.62\pm0.20)\times10^{-9}$ & -\\
   2005$-$489 & SS & 2020-02-18 & $7.4$ & $2.6$ & $3$ & $751.7$ & $267.7$ & $5.4$ & $235.1\pm23.5$ & $1.4\pm0.5$ & $(2.9\pm2.1)\times10^{10}$ & $(2.90\pm0.09)\times10^{-9}$ & 0.03\\
   2155$-$304 & SS & 2020-02-18 & $12.3$ & $2.6$ & $-1$ & $320.3$ & $161.2$ & $2.0$ & $186.0\pm18.6$ & $3.0\pm0.5$ & $(5.5\pm2.0)\times10^{9}$ & $(2.00\pm0.02)\times10^{-8}$ & 0.15 \\
   2322$-$409 & SS & 2020-11-20 & $10.0$ & $3.5$ & $0$ & $55.8$ & $36.9$ & $0.6$ & $43.5\pm4.4$ & $2.3\pm0.7$ & $(2.3\pm1.4)\times10^{9}$ & $(1.07\pm0.06)\times10^{-9}$ & 0.011\\
   2356$-$309 & SS & 2021-12-04 & $16.6$ & $3.2$ & $1$ & $24.1$ & $12.5$ & $0.3$ & $14.0\pm1.4$ & $2.6\pm0.6$ & $(5.8\pm2.9)\times10^{8}$ & $(4.50\pm0.39)\times10^{-10}$ & 0.02\\
     \hline
    \end{tabular}}
    \tablefoot{
    \tablefoottext{a}{Source morphology: C - core only; SS - single-sided; DS - double-sided jet morphology.}
    \tablefoottext{b}{Beam major axis.}
    \tablefoottext{c}{Beam minor axis.}
    \tablefoottext{d}{Beam position angle.}
    \tablefoottext{e}{Total flux density.}
    \tablefoottext{f}{Rms noise of the image.}
    \tablefoottext{g}{Lowest contour.}
    \tablefoottext{h}{Flux density of the core.}
    \tablefoottext{i}{FWHM of the core.}
    \tablefoottext{j}{Brightness temperature.}
    \tablefoottext{k}{Photon flux from \citep{abdollahi22}.}
    \tablefoottext{l}{Very high energy flux in Crab units.}
    }
\end{sidewaystable*}

\end{appendix}

\end{document}